\documentclass[conference]{IEEEtran}
\IEEEoverridecommandlockouts
\usepackage{cite}
\usepackage{amsmath,amssymb,amsfonts}
\usepackage{algorithmic}
\usepackage{graphicx}
\usepackage{textcomp}
\usepackage{xcolor}
\def\BibTeX{{\rm B\kern-.05em{\sc i\kern-.025em b}\kern-.08em
    T\kern-.1667em\lower.7ex\hbox{E}\kern-.125emX}}
    
\usepackage{flushend}
\usepackage{epsfig}
\usepackage{cite}
\usepackage{amsmath}
\usepackage{flushend}
\usepackage{multirow}
\usepackage{color}
\usepackage{enumerate}
\usepackage{easylist}
\usepackage{amsthm}
\usepackage{amssymb }
\usepackage{bm}
\usepackage{cite}
\usepackage{float}
\usepackage{mathrsfs}
\usepackage{amsfonts}
\usepackage{bm}
\usepackage{cite}
\usepackage{float}
\usepackage{mathrsfs}
\usepackage{graphicx}
\usepackage{subfigure}
\usepackage{tabularx}
\def\S{S}
\def\W{\mathit{W}}
\def\V{\bm{V}}
\def\Q{\bm{Q}}
\def\K{\bm{K}}
\def\Z{\bm{Z}}

\def\E{\bm{E}}

\def\N{N}
\def\p{p}
\def\R{\mathbb{R}}
\def\Nd{^{\N\times\d}}
\def\d{d}

\def\st{\S^{2}}
\def\std{^{\st\times\d}}

\def\dh{d_h}
\def\l{l}
\def\h{h}

\def\L{L}
\def\j{_{j}}
\def\x{{\mathbf x}}
\def\z{{\mathbf z}}
\def\xp{\x^{\p}}

\def\yb{\mathbf{y}}

\begin{document}

\title{TEDGE-Caching: Transformer-based Edge Caching Towards 6G Networks\\
{}
\thanks{This Project was partially supported by Department of National Defence's Innovation for Defence Excellence \& Security (IDEaS), Canada.}
}

\author{\IEEEauthorblockN{1\textsuperscript{st} Zohreh Hajiakhondi Meybodi}
\IEEEauthorblockA{\textit{Electrical and Computer Engineering} \\
\textit{Concordia University}\\
Montreal, Canada }
\and
\IEEEauthorblockN{2\textsuperscript{nd} Arash Mohammadi}
\IEEEauthorblockA{\textit{Concordia Ins. for Inf. Systems Eng.} \\
\textit{Concordia University}\\
Montreal, Canada }
\and
\IEEEauthorblockN{3\textsuperscript{nd} Elahe Rahimian}
\IEEEauthorblockA{\textit{Concordia Ins. for Inf. Systems Eng.} \\
\textit{Concordia University}\\
Montreal, Canada }
\and
\IEEEauthorblockN{4\textsuperscript{nd} Shahin Heidarian}
\IEEEauthorblockA{\textit{Electrical and Computer Engineering} \\
\textit{Concordia University}\\
Montreal, Canada }
\and
\IEEEauthorblockN{5\textsuperscript{nd} Jamshid Abouei}
\IEEEauthorblockA{\textit{Department of Electrical Engineering} \\
\textit{Yazd University}\\
Yazd, Iran }
\and
\IEEEauthorblockN{6\textsuperscript{nd} Konstantinos~N.~Plataniotis}
\IEEEauthorblockA{\textit{Electrical \& Computer Engineering} \\
\textit{University of Toronto,}\\
Toronto, Canada }
}

\maketitle

\begin{abstract}
As a consequence of the COVID-19 pandemic, the demand for telecommunication for remote learning/working and telemedicine  has significantly increased. Mobile Edge Caching (MEC) in the 6G networks has been evolved as an efficient solution to meet the phenomenal growth of the global mobile data traffic by bringing multimedia content closer to the users. Although massive connectivity enabled by MEC networks will significantly increase the quality of communications, there are several key challenges ahead. The limited storage of edge nodes, the large size of multimedia content, and the time-variant users' preferences make it critical to efficiently and dynamically predict the popularity of content to store the most upcoming requested ones before being requested. Recent advancements in Deep Neural Networks (DNNs) have drawn much research attention to predict the content popularity in proactive caching schemes. Existing DNN models in this context, however, suffer from long-term dependencies, computational complexity, and unsuitability for parallel computing. To tackle these challenges, we propose an edge caching framework incorporated with the attention-based Vision Transformer (ViT) neural network, referred to as the Transformer-based Edge (TEDGE) caching, which to the best of our knowledge, is being studied for the first time. Moreover, the TEDGE caching framework requires no data pre-processing and additional contextual information. Simulation results corroborate the effectiveness of the proposed TEDGE caching framework in comparison to its counterparts.
\end{abstract}


%OOOOOOOOOOOOOOOOOOOOOOOOOOOOOOOOOOOOOOOOOOOOOOOOOOOOOOOO
\section{Introduction} \label{Sec:1}
%OOOOOOOOOOOOOOOOOOOOOOOOOOOOOOOOOOOOOOOOOOOOOOOOOOOOOOOO
Mobile Edge Caching (MEC)~\cite{Tang2020, Fadlullah2020} is an emerging technology in the Beyond  Fifth Generation (5G) communication networks (also referred to as 6G) developed to meet the phenomenal growth of the global mobile data traffic. Enabling caching at the edge of the network provides the opportunity to store popular content at the storage of the heterogeneous next generation Node B (hgNB) during the off-peak intervals~\cite{Vallero2020,Chen2020}. After requesting a content by an edge (e.g., Internet of Thing (IoT)) device, this request is directly served by the neighboring hgNBs, having the requested content. In such scenarios, the cache-hit occurs; otherwise, it is known as a cache-miss and the requested content is sent from the content server to the hgNB to serve the request~\cite{Hajiakhondi2019, Hajiakhondi2020}. Integrating Unmanned Aerial Vehicles (UAVs) as the flying hgNBs into the terrestrial MEC networks~\cite{Hajiakhondi2021-2} extends the service coverage and improves the Quality of Service (QoS) of User Equipment (UE) in the Beyond 5G networks. Due to the limited local storage capacity of cache-enabled hgNBs, it is of significant importance to identify/store the most popular content to enhance the cache efficiency of the network. In the MEC networks, there are two types of caching strategies, i.e., Reactive caching and proactive caching. Conventional reactive caching schemes~\cite{Giovanidis2016}, such as First-In-First-Out (FIFO), Least Recently Used (LRU), and Least Frequently Used (LFU) frameworks, identify the most popular content based on the underlying pattern of observed users' requests. A critical drawback of reactive caching is that popular content can only be identified after being requested. As a consequence, they are not robust to the dynamically changing behavior of the content popularity. Therefore, the main focus of recent researches has been shifted to use proactive caching, e.g., using Deep Neural Networks (DNN) models to predict the Content Popularity (CP) from the request patterns. In this context, popular content can be  dynamically allocated in the storage of hgNBs before being requested. The paper aims to further advance this emerging~field.

%OOOOOOOOOOOOOOOOOOOOOOOOOOOOOOOOOOOOOOOOOOOOOOOOOOOOOOOOO
\vspace{.1in}
\noindent
\textbf{Literature Review:}
%OOOOOOOOOOOOOOOOOOOOOOOOOOOOOOOOOOOOOOOOOOOOOOOOOOOOOOOOO
Generally speaking, both temporal and spatial correlations exist within the  time-variant request pattern of multimedia content. While spatial correlation reflects different users' preferences, depending on the geographical location and users' contextual information, the temporal correlation represents the variation of content popularity over time. In this context, several DNN models~\cite{Doan2018, Ale2019, Fan2021, Zhang2019, Rathore2019, Lin2020, Zhong2020, Wu2019, Wang2019} are introduced to capture the temporal and/or spatial features of user preferences in proactive caching schemes. For instance, Yu~\textit{et al.}~\cite{Yu2021} used an auto-encoder model to predict users' preferences in the future by learning the latent representation of raw data in an unsupervised fashion.  Auto-encoder models, however, suffer from training complexity. Tsai~\textit{et al.}~\cite{Tsai2018} used Convolutional Neural Network (CNN) for predicting users' interests based on sentence analysis. Ndikumana~\textit{et al.}~\cite{Ndikumana2021} introduced a DNN-based caching framework, compromising of Multi-Layer Perceptron (MLP) and CNN models, where contextual information such as age, emotion, and gender are utilized for making caching decisions. Although CNN-based proactive caching schemes have local spatial feature awareness, they are inefficient for extracting temporal features from the patterns of sequential requests. Furthermore, such models require multi-source input such as regional information, and contextual information of users to improve the cache performance. Therefore, they need an efficient data pre-processing model to extract this information. 

To deal with the time-varying behavior of request patterns, Recurrent Neural Networks (RNNs), such as Long Short Term Memory (LSTM)~\cite{Zhang2019, Mou2019}, are introduced to use historical information of the content. To extract both spatial and temporal features of CP data, Ale~\textit{et al.}~\cite{Ale2019} used a combination of LSTM and CNN models. LSTM-based caching frameworks, however, suffer from long-term dependencies, computation complexity, and unsuitability for parallel computing. To address challenges associated with RNN architectures, the Transformer neural network~\cite{Vaswani2017} has been designed to handle sequential input data, which is purely reliant on attention mechanisms with no recurrence or convolutions. One of the most important advantages of Transformers over RNN models is the attention mechanism, which eliminates the need to analyze data in the same order. Consequently, Transformers have higher parallelization capabilities than RNNs, implying reduced training time. 

%OOOOOOOOOOOOOOOOOOOOOOOOOOOOOOOOOOOOOOOOOOOOOOOOOOOOOOOOO
\vspace{.1in}
\noindent
\textbf{Contributions:} Motivated by the above discussion, we introduce a Vision Transformer-based Edge (TEDGE) caching framework with the application to the MEC networks. The proposed TEDGE  framework learns the real-time caching strategy from sequential requests of multimedia content. The main objective of several recent time-series prediction models that have been applied to the multimedia content caching~\cite{Zhang2021, Zhang2021_2} is to predict the underlying patterns of the future multimedia content requests, i.e., the number of content requests using historical  information. Considering the fact that the users' preferences remain unchanged for a while~\cite{Lin2020}, it is sufficient to predict the potential Top-$K$ popular content using the learned patterns from historical requests. The main focus of this study, therefore, is to predict the Top-$K$ popular content using historical information instead of predicting the number of upcoming requests. In summary, the paper makes the following key contributions:
\begin{itemize}
\item The TEDGE caching framework is an edge-assisted intelligent caching framework
that learns the caching strategy from the historical request patterns without relying on data pre-processing or feature engineering. More precisely, the TEDGE caching framework is a multi-label classification model with the aim of minimizing the difference between the actual Top-$K$ popular content and the predicted ones.
\item To simultaneously analyze the sequential pattern of all content, the TEDGE caching framework employs a ViT architecture instead of using conventional Transformer models. The input of the ViT model is an image, where each pixel indicates the number of requests of each content in a specific time. 
\end{itemize}
 Simulation results based on real-trace of multimedia requests illustrate that the proposed TEDGE caching framework outperforms its state-of-the-art counterparts in cache-hit-ratio. The remainder of the paper is organized as follows: In Section~\ref{Sec:2}, the system model is described and the main assumptions required for  implementation of the proposed TEDGE framework are introduced. Section~\ref{Sec:3} presents the proposed TEDGE caching framework. Simulation results are presented in Section~\ref{Sec:4}. Finally, Section~\ref{Sec:5} concludes the paper.

%OOOOOOOOOOOOOOOOOOOOOOOOOOOOOOOOOOOOOOOOOOOOOOOOOOOOOOOO
\section{System Model and Problem Description} \label{Sec:2}
%OOOOOOOOOOOOOOOOOOOOOOOOOOOOOOOOOOOOOOOOOOOOOOOOOOOOOOOO
%%%%%%%%%%%%%%%%%%%%%%%%%%%%%%%%%%%%%%%%%%
\begin{figure} [!t]
\centering \includegraphics [scale = 0.2] {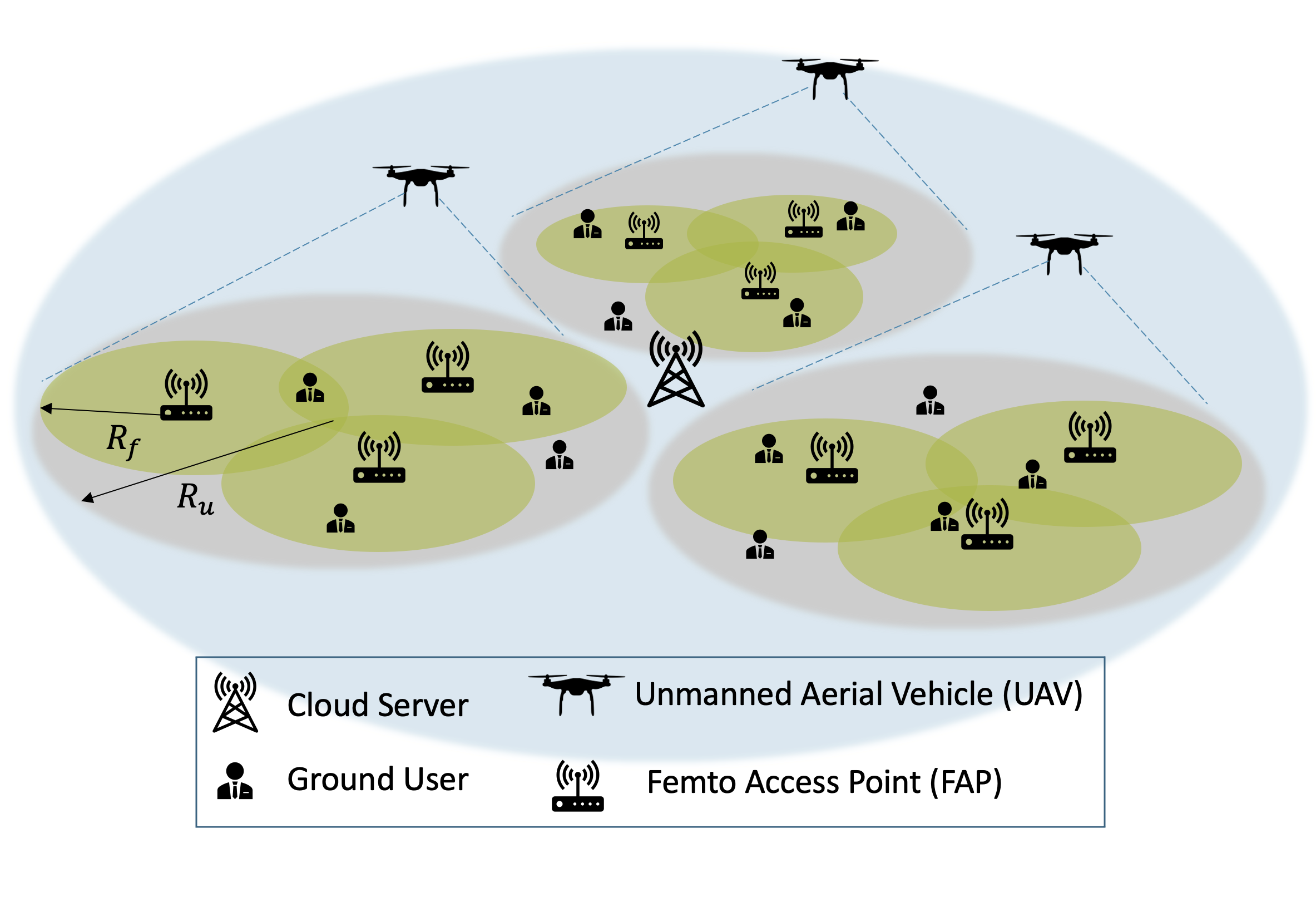}
\caption{\footnotesize A typical structure of the UAV-aided cellular network.} \label{sys}
\vspace{-.1in}
\end{figure}
%%%%%%%%%%%%%%%%%%%%%%%%%%%%%%%%%%%%%%%%%%
We consider a UAV-aided cellular network with heterogeneous Radio Access Technologies (RATs) as the $6$G network model. There are $N_h$ number of hgNBs, consisting of $N_u$ number of UAVs, denoted by $u_{k}$, for ($1 \leq k \leq N_u$), along with $N_f$ number of FAPs, denoted by $f_i$, for ($1 \leq i \leq N_f$), serving requests of $N_g$ number of UEs, denoted by $U_j$, for ($1 \leq j \leq N_g$). The hgNBs are equipped with a limited cache size, denoted by $K$. As shown in Fig.~\ref{sys}, FAPs are independently and randomly distributed in the environment following a Poisson Point Process (PPP)~\cite{Chen2017_2}. We also consider a Gaussian mixture distribution for UEs, leading to a dense population in some environments. Due to the movement of UEs, the population is changed over time, therefore, the location of UAVs are determined by the $K$-means clustering algorithm~\cite{Hajiakhondi2021}, where UAVs remain hovering at their locations while serving a request for data delivery~\cite{Fadlullah2020}. A Software Defined Network (SDN) controller is used to manage the aerial and terrestrial connections and control the link quality and topology of UAVs~\cite{Fadlullah2020}. We denote a library of content $\mathcal{C} = \lbrace c_1, \ldots, c_{N_c} \rbrace$, where  $N_c=|\mathcal{C}|$ is the cardinality of contents in the network. For simplicity, it is assumed that the size of all contents $c_l$, for ($1 \leq l \leq N_c$), are the same~\cite{Fadlullah2020}, and UEs request at most one content in each time slot. 

CP in multimedia services follows Mandelbrot–Zipf (M-Zipf) distribution~\cite{Wang2020}, where the global probability of requesting content $c_l$ by all UEs, denoted by $p_l$, is given by
\begin{equation}\label{nEq:5}
p_l=\dfrac{(l+\zeta)^{-\gamma}}{\sum \limits_{r=1}^{N_c}{(r+\zeta)^{-\gamma}}},
\end{equation}
where $\gamma$ and $\zeta$ represent the skewness and plateau factors, respectively, and term $r$ is the rank of content $c_r$, when all contents are sorted in descending order of their popularity. In addition, $p_l^{(b)}$ represents the local probability of requesting content $c_l$ by IoT devices located in the coverage area of hgNB $b$, where $b$ could be $u_k$, for ($1 \leq k \leq N_u$), or $f_i$, for ($1 \leq i \leq N_f$), where $p_l=\sum\limits_{b=1}^{N_h}p_l^{(b)}$.

%OOOOOOOOOOOOOOOOOOOOOOOOOOOOOOOOOOOOOOOOOOOOOOOOOOOOOOOO
\section{TEDGE Caching Framework} \label{Sec:3}
%OOOOOOOOOOOOOOOOOOOOOOOOOOOOOOOOOOOOOOOOOOOOOOOOOOOOOOOO
In this section, we present the TEDGE caching framework, which is designed to predict the Top-$K$ popular content. To be specific, we first briefly introduce the dataset used in this study, and also present the preparation phase to adopt the dataset to the TEDGE caching framework. Then, we explain different blocks of the ViT architecture, which is used as the multi-label classification within the TEDGE caching framework.

%OOOOOOOOOOOOOOOOOOOOOOOOOOOOOOOOOOOOOOOOOOOOOOOOOOOOOOOO
\subsection{Dataset} 
%OOOOOOOOOOOOOOOOOOOOOOOOOOOOOOOOOOOOOOOOOOOOOOOOOOOOOOOO
In this study, we use MovieLens Dataset~\cite{Harper2015}, which is one of the well-known movie recommendation services. In this dataset, movies with related information such as movie titles, release date, and genre are provided. Each content is requested by several users in different timestamps, where the contextual information of users such as age, gender, occupation, and their ZIP codes are also released. With the assumption that users leave a comment after watching a movie~\cite{Zhang2019, Dernbach2016, Li2016} and in order to extract the content request pattern, commenting on a content is considered as a request. Moreover, to identify the users' location in each timestamp, ZIP codes are converted to longitude and latitude coordinates~\cite{Zhang2019}. Considering the limited transmission range of hgNBs, hgNBs' locations, and users locations, the available hgNBs for serving requests of all users will be determined. Our main goal in the TEDGE caching framework is to monitor the historical requests pattern of each content to predict the Top-$K$ popular content in an upcoming time period. Therefore, the preparation of the dataset is performed in the following four steps: 

\vspace{.1in}
\noindent
\textit{\textbf{Step 1 (Request Matrix Formation)}}: In the first step, the dataset is sorted for each content $c_l$, for ($1 \leq l \leq N_c$), in the ascending order of time. Therefore, we form an ($T \times N_c$) indicator request matrix for each hgNB, denoted by $\textbf{R}$, where $T$ and $N_c$ represent the total number of timestamps and the total number of distinct content, respectively. In the request matrix, $r_{t,l}=1$ illustrates that content $c_l$ is requested at time $t$; otherwise, $r_{t,l}=0$.

\vspace{.1in}
\noindent
\textit{\textbf{Step 2 (Time Windowing)}}: Considering the fact that the most popular content should be cached at the storage of hgNBs during the off-peak time~\cite{Vallero2020}, there is no need to predict the content popularity at each timestamp. We, therefore, define the updating time $t_u$ (i.e., the off-peak time), as the timestamp that the storage of hgNBs is updated by the new popular content. In this case, we will have a time window with the length of $\mathcal{W}$, where $\mathcal{W}$ is associated to the time duration between two updating times, and the number of time windows is represented by $N_{\mathcal{W}}=\frac{T}{\mathcal{W}}$. Therefore, we have a ($N_{\mathcal{W}} \times N_c$) window-based request matrix, denoted by $\textbf{R}^{(\mathcal{W})}$, where $r^{(w)}_{t_u,l}= \sum \limits_{t=(t_u-1)\mathcal{W}+1}^{t_u\mathcal{W}} r_{t,l}$ illustrates the total number of requests of content $c_l$ between updating time $t_u-1$ and $t_u$.

\begin{figure}[t!]
\centering
\includegraphics[scale=0.37]{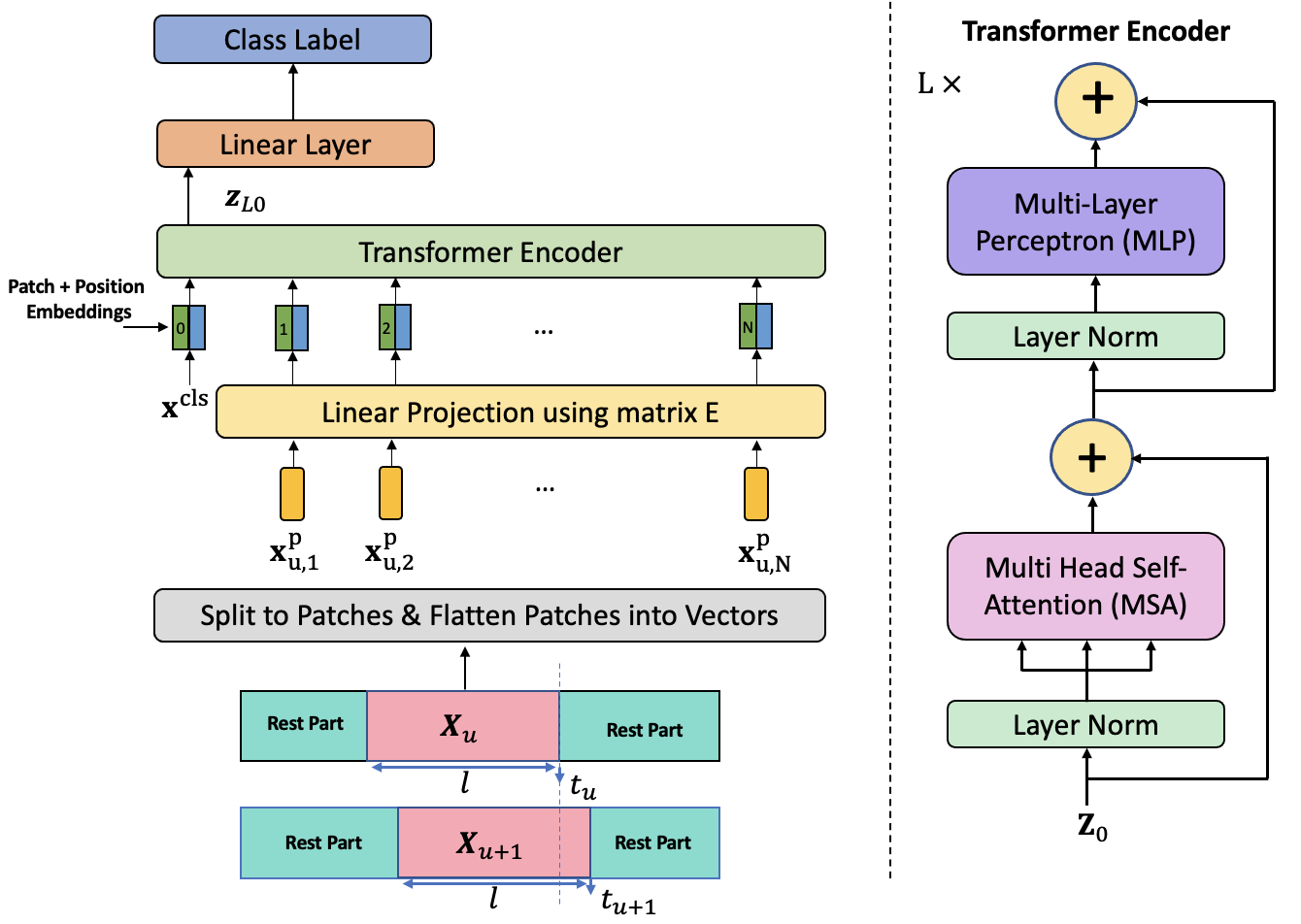}
\caption{Left: Pipeline of the ViT Architecture, Right: Architecture of the Transformer Encoder.}\label{BlockDiagram}
\end{figure}

\vspace{.1in}
\noindent
\textit{\textbf{Step 3 (Data Segmentation)}}: As mentioned previously, the main target of the TEDGE caching framework is to use the historical information of content to predict the Top-$K$ popular content in the next updating time. Given the window-based request matrix $\textbf{R}^{(\mathcal{W})}$, the collected request pattern data is segmented via an overlapping sliding
window of length $l$. As it can be seen from Fig.~\ref{BlockDiagram}, the window-based request matrix $\textbf{R}^{(\mathcal{W})}$ is converted into $\mathcal{D}=\{(\textbf{X}_u, \textbf{y}_u)\}_{u=1}^{M}$, where $M$ represents the total number of segments. Moreover, terms $\textbf{X}_u \in \mathbb{R}^{M\times N_c}$ and $ \textbf{y}_u \in \mathbb{R}^{N_c \times 1}$ represent the request pattern of all content before updating time $t_u$ with the length of $l$, and its corresponding label, respectively. Considering the fact that there are $N_c$ number of content through the network, and our objective is to predict the Top-$K$ popular content in the next updating time, the problem at hand is a multi-label classification, where ${y_u}_l=1$ illustrates that content $c_l$ would be popular at $t_{u+1}$. Therefore, $c_l$ should be stored at the storage of hgNB to increase the cache-hit-ratio.
%%%%%%%%%%%%%%%%%%%%%%%%%%%%%%%%%%%%%%%%
\setlength{\textfloatsep}{0pt}
%%%%%%%%%%%%%%%%%%%%%%%%%%%%%%%%%%%%%%%%

\vspace{.1in}
\noindent
\textit{\textbf{Step 4 (Data Labeling)}}: Due to the limited storage of hgNBs, it is sufficient to identify the Top-$K$ popular content, instead of predicting the popularity of all content at each updating time. According to the request pattern of multimedia content, we calculate the probability of requesting content $c_l$, for ($1 \leq l \leq N_c$), which is obtained as follows
\begin{equation}
p_l^{(t,b)}=\dfrac{r^{(w)}_{t_u,l}}{\sum \limits_{l=1}^{N_c}r^{(w)}_{t_u,l}}.
\end{equation}
Note that, relying on the probability of content as a single criteria for identifying the popularity of content has the following disadvantages: (i) Popular content with a high number of requests will be identified as the Top-$K$ popular content for a long time, even if they are becoming unpopular, and; (ii) The popularity of new/unknown coming content (first appearance) would be predicted with a considerable delay, because the cumulative number of requests of such content is less than other content that are existing for a long time. To tackle with this issue, we use the skewness of the request pattern as another metric, which is a widely used indicator in time-series forecasting models~\cite{Joseph2001}. The skewness of content $c_l$ is denoted by $\zeta_l$,  where $\zeta_l < 0$ shows that the number of requests of content $c_l$ increases over time. Finally, the Top-$K$ content with the highest probability and the negative skew will be labeled as the Top-$K$ popular content. This completes presentation of the data preparation for training the TEDGE caching framework. Next, we present the ViT architecture.

%=========================================================
\subsection{ViT Architecture}
%=========================================================
Generally speaking, the main characteristics of Transformers are as follows: (i) \textit{Non-sequential:} Unlike RNN, the Transformer's attention mechanism makes it unnecessary to process data in the same order. As a result, Transformer is more parallelization than RNNs, which means it takes less time to train; (ii) \textit{Self Attention}, indicating the similarity scores between different elements of a sequential data, and; (iii) \textit{Positional Embeddings:} Since Transformers are non-sequential learning models, the order of information in a sequential data is missing. Therefore, Positional embeddings is introduced for recovering position information. As it can be seen from Fig.~\ref{BlockDiagram}, the TEDGE caching framework consists of the following three modules: (a) Patch and Position Embeddings; (b) Transformer encoder, and; (c) MLP head, which are described as follows:

\vspace{.1in}
\noindent
\textit{\textbf{Patch and Position Embeddings:}} As it can be seen from Fig.~\ref{BlockDiagram}, the segmented CP data $\textbf{X}_u$ is split into $\N$ sequence of non-overlapping patches with the fixed-size of $(S\times S)$, where the total number of patches is $N = w / S$. After this step, each patch is flattened into a vector $\xp_{u,\j} \in \R^{\st}$ for ($1 \leq j \leq \N$). To embed vector $\xp_{u,\j} \in \R^{\st}$ into the model's dimension $\d$, a linear projection  $\E\in\R\std$ is used, which is shared among all patches, where the output of this projection is referred to as the patch embeddings. We append a learnable embedding token $\x^{cls}$ to the beginning of the sequence of embedded patches~\cite{Bert}. Finally, the position embeddings $\E^{pos}\in\R^{(\N + 1)\times\d}$, is added to the patch embeddings to explicitly encode the order of the input sequence. The output of the patch and position embeddings $\Z_0$ is given by
\begin{eqnarray}
\Z_0 = [\x^{cls}; \xp_{u,1}\E; \xp_{u,2}\E;\dots; \xp_{u,\N}\E] + \E^{pos}. \label{eq:patch}
\end{eqnarray}
%

%\vspace{.1in}
\noindent
\textit{\textbf{Transformer Encoder:}} Given the output of the linear projection, the sequence of vectors $\Z_0$ is fed to the transformer encoder~\cite{Vaswani2017}. As it can be seen from Fig.~\ref{BlockDiagram}, the transformer encoder consists of $\L$ layers, with two modules, i.e., the Multihead Self-Attention (MSA) mechanism, and the MLP modules, where MLP module consists of two linear layers with Gaussian Error Linear Unit (GELU) activation function. The output of the MSA and MLP modules of layer $l$, for ($1 \leq l \leq L$) are given by 
\begin{eqnarray}
\Z^{'}_l &=& MSA(LayerNorm(\Z_{\l-1})) + \Z_{\l-1},\label{eq:MSA}\\
\Z_l &=& MLP(LayerNorm(\Z^{'}_{\l})) + \Z^{'}_{\l}, \label{eq:MLP}
\end{eqnarray}
where a layer-normalization is used to avoid the degradation problem~\cite{layernorm}. Finally, the output of the Transformer is 
\begin{eqnarray}
\Z_\L = [\z_{\L0}; \z_{\L1}; \dots; \z_{\L\N}],
\end{eqnarray}
where $\z_{\L0}$ is used for classification purposes, which is passed to a Linear Layer (LL), i.e.,
\begin{eqnarray}
\yb = LL(LayerNorm(\z_{\L0})).\label{eq:out}
\end{eqnarray}
This completes the description of the Transformer autoencoder. Next, we present the description of the SA and the MSA, respectively.

%------------------------------------------------------------------------------------------------------------
\vspace{.1in}
\noindent
\textit{\textbf{1. Self-Attention (SA):}} The SA module~\cite{Vaswani2017} is used in the Transformer architecture to focus on significant parts of a given input by capturing the interaction between different vectors in $\Z \in \R\Nd$, where $\Z$ consists of $\N$ vectors, each with an embedding dimension of $\d$. Towards this goal, three different matrices are defined, named Queries $\Q$, Keys $\K$, and Values $\V$, computed by a linear transformation as follows
\begin{eqnarray}
[\Q, \K, \V] = \Z\W^{QKV}\label{eq.2},
\end{eqnarray}
where $\W^{QKV} \in \R^{\d\times 3\dh}$ represents the trainable weight matrix, and $\dh$ is the dimension of $\Q$, $\K$, and $\V$. The SA block measures the pairwise similarity between each query and all keys. The output of the SA block $SA(\Z) \in \R^{\N\times \dh}$, which is the weighted sum over all values $\V$, is given by
\begin{eqnarray}
SA(\Z) = \text{softmax}(\frac{\Q\K^T}{\sqrt{\dh}}) \V\label{eq.3}, 
\end{eqnarray}
where term $\dfrac{\Q\K^T}{\sqrt{\dh}}$ is the scaled dot-product of $\Q$ and $\K$ by $\sqrt{\dh}$ and $\text{softmax}$ is used to convert the scaled similarity to the probability. 
%%%%%%%%%%%%%%%%%%%%%%%%%%%%%%%%%%%%%%%%
\begin{table*}[t]
\centering
\renewcommand\arraystretch{2}
\caption{\small Variants of the TEDGE caching framework.}
\label{table1}
{\begin{tabular}{   c | c c c c c c | c  c c}
\hline
\hline
\textbf{Model ID}
& \textbf{Layers}
& \textbf{Model dimension $\d$}
& \textbf{MLP layers}
& \textbf{MLP size}
& \textbf{Heads}
& \textbf{Params}
& \textbf{Accuracy}
& \textbf{Loss}
\\
\hline
     \textbf{1}
& 1
& 32
& 1
& 256
& 8
& 887,140
& 91.09 $\%$
& 0.2023 
\\
    \textbf{2}
& 1
& 64
& 1
& 256
& 8
& 1,822,055
& 91.70 $\%$
& 0.1888 
\\
\textbf{3}
& 1
& 128
& 1
& 256
& 8
& 3,913,063
& 92.61 $\%$
& 0.1645
\\
 \textbf{4}
& 2
& 128
& 1
& 256
& 8
& 4,506,983
& 91.68 $\%$
& 0.1961 
\\
\textbf{5}
& 1
& 128
& 2
& 256
& 8
& 3,978,852
& 93.19 $\%$
& 0.1495
\\
\textbf{6}
& 1
& 128
& 3
& 256
& 6
& 3,912,807
& 92.89 $\%$
& 0.1600
\\
\textbf{7}
& 1
& 128
& 3
& 256
& 8
& 4,044,644
& \textbf{93.72} $\%$
& 0.1391
\\
\textbf{8}
& 1
& 128
& 3
& 256
& 10
& 4,176,487
& 92.61 $\%$
& 0.1650
\\
\textbf{9}
& 1
& 128
& 1
& 512
& 8
& 7,215,719
& 91.48 $\%$
& 0.2060 
\\
\hline
\end{tabular}}
\end{table*}
%%%%%%%%%%%%%%%%%%%%%%%%%%%%%%%%%%%%%%%%

%------------------------------------------------------------------------------------------------------------
\vspace{.1in}
\noindent
\textit{\textbf{2. Multihead Self-Attention (MSA):}} In the MSA, the SA block is performed $\h$ times in parallel, which results in attending to information from different representation subspaces at different positions. More precisely, MSA module consists of $\h$ heads with different trainable weight matrix for each head $\{\W^{QKV}_i\}^{\h}_{i=1}$. After applying the SA mechanism on input $\Z$ for each head (Eqs.~\eqref{eq.2}-\eqref{eq.3}), the outputs of $\h$ heads are concatenated into a single matrix $[SA_1(\Z); SA_2(\Z); \dots; SA_h(\Z)]\in \R^{\N\times \h.\dh}$. Finally, the output of the MSA module is given by
\begin{eqnarray}
MSA(\Z) = [SA_1(\Z); SA_2(\Z); \dots; SA_h(\Z)]W^{MSA},
\end{eqnarray}
where $\W^{MSA} \in \R^{\h\dh \times \d}$ and $\dh$ is set to $\d / \h$.

%OOOOOOOOOOOOOOOOOOOOOOOOOOOOOOOOOOOOOOOOOOOOOOOOOOOOOOOO
\section{Simulation Results} \label{Sec:4}
%OOOOOOOOOOOOOOOOOOOOOOOOOOOOOOOOOOOOOOOOOOOOOOOOOOOOOOOO
In this Section,  we first evaluate different variants of the proposed TEDGE caching framework to obtain the best architecture through the process of trial and error.  Considering the location of UE, which is obtained from the ZIP code and followed by Reference~\cite{Ndikumana2021}, six hgNBs are employed in different areas, where the classification accuracy is averaged over all hgNBs. In all experiments, the one-dimensional time-series content's request data is converted to a sequential set of images, which is known as the Gramian Angular Field (GAF) technique~\cite{Hong2020}. Utilizing GAF method, not only the temporal characteristics of the  data is preserved, but also the temporal correlations of data are included. In Back Propagation (BP) training, Adam optimizer is employed, where the weight decay and betas are set to $0.001$ and ($0.9, 0.999$). The size of the input image, the size of input patches, and the batch size are ($25 \times 25$), ($5\times5$), and $256$, respectively. Finally, we use binary cross-entropy as the loss function for our  multi-label classification problem. According to the results in Table~\ref{table1}, increasing the model dimension from $32$ to $128$ (Model $1$ to Model $3$), and also the number of MLP layers from $1$ to $3$ (Models $3$, $5$, and $6$) increase the classification accuracy, while increasing the number of trainable parameters. Moreover, we evaluate the effect of MLP size on the classification accuracy. As it can be seen from  Table~\ref{table1}, increasing the MLP size from $256$ to $512$ (Model $3$ and Model $9$) decreases the classification accuracy. Similarly, there is no improvement in the classification accuracy by increasing the number of transformer layers (see Models $3$ and $4$). Furthermore, considering Model $6$ to Model $8$, the best number of heads in this architecture is equal to $8$. Note that we also evaluated the effect of window length on the classification accuracy, while no improvement has been achieved by changing the window length.

\begin{figure}[t!]
\centering
\includegraphics[scale=0.35]{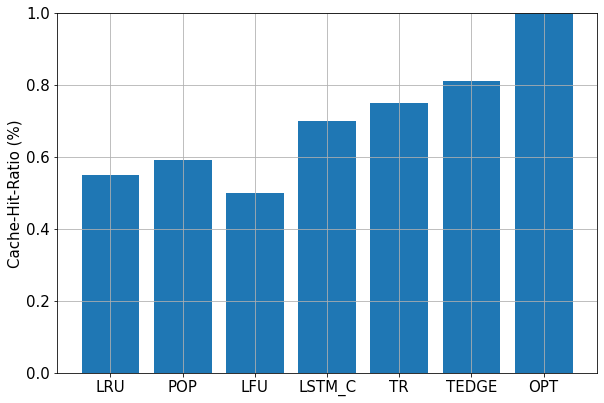}
\caption{A comparison with state-of-the-arts based on the cache-hit-ratio. }\label{cachehit}
\end{figure}

Finally, we compare the performance of the proposed TEDGE caching framework with five state-of-the-art caching schemes on the Movielens dataset, including LRU, LFU, PopCaching~\cite{Li2016}, LSTM-C~\cite{Zhang2019}, and the TRansformer (TR) caching, which is an upgraded version of the attention-based neural network in Reference~\cite{Nguyen2019}. While the attention-based model in Reference~\cite{Nguyen2019} is used for predicting the request pattern of online content, we adopt it to predict the Top-$K$ popular ones. Fig.~\ref{cachehit} compares the performance of the proposed TEDGE scheme with other baselines mentioned above from the aspect of the cache-hit ratio, when the DNN models reach the steady state. In the content caching context, cache-hit-ratio is a widely used metric, illustrating the ratio of requests served by hgNBs versus total requests. Considering the Zipf distribution for the content popularity profile, we set the storage capacity of hgNBs to $10\%$ of the total content~\cite{Hajiakhondi2019}. As shown in Fig.~\ref{cachehit}, the optimal strategy~\cite{Zhang2019} is a caching scheme, where all requests are served through hgNBs, which cannot be obtained in reality. According to the results in Fig.~\ref{cachehit}, the proposed TEDGE caching framework obtains the highest cache-hit-ratio in comparison to its state-of-the-art counterparts.

%OOOOOOOOOOOOOOOOOOOOOOOOOOOOOOOOOOOOOOOOOOOOOOOOOOOOOOOO
\section{Conclusion} \label{Sec:5}
%OOOOOOOOOOOOOOOOOOOOOOOOOOOOOOOOOOOOOOOOOOOOOOOOOOOOOOOO
%%%%%%%%%%%%%%%%%%%%%%%%%%%%%%%%%%%%%%%%
\setlength{\textfloatsep}{0pt}
%%%%%%%%%%%%%%%%%%%%%%%%%%%%%%%%%%%%%%%%
In this paper, we presented a Transformer-based Edge (TEDGE) caching framework with the application to the Mobile Edge Caching (MEC) networks. In order to efficiently learn the real-time caching strategy from the time-series request pattern of multimedia content, we employed a Vision Transformer (ViT) architecture. To the best of our knowledge, this is the first time that a ViT architecture is used in MEC networks to increase the cache-hit-ratio by simultaneously identifying the Top-$K$ popular content with high accuracy. Simulation results showed that the proposed TEDGE caching-CS scheme improves the cache-hit ratio when compared to its state-of-the-art counterparts.


\begin{thebibliography}{00}
\bibitem{Tang2020}
F. Tang, Y. Kawwamoto, N. Kato, and J. Liu, \newblock ``Edge Cloud Server Deployment with Transmission Power Control through Machine Learning for 6G Internet of Things,'' \newblock {\em IEEE Trans. Emerg. Topics Comput.}, Dec. 2019.

\bibitem{Fadlullah2020}
Z. M. Fadlullah and N. Kato, \newblock ``HCP: Heterogeneous Computing Platform for Federated Learning Based Collaborative Content Caching Towards 6G Networks,''
\newblock {\em IEEE Trans. Emerg. Topics Comput.}, Apr. 2020.

\bibitem{Vallero2020}
G. Vallero, M. Deruyck, W. Joseph and M. Meo, \newblock ``Caching at the edge in high energy-efficient wireless access networks,''
\newblock {\em IEEE International Conference on Communications (ICC)}, June 2020, pp. 1-7.

\bibitem{Chen2020}
J. Chen, H. Xing, X. Lin and S. Bi, \newblock ``Joint Cache Placement and Bandwidth Allocation for FDMA-based Mobile Edge Computing Systems,''
\newblock {\em IEEE International Conference on Communications (ICC)}, June 2020, pp. 1-7.

\bibitem{Hajiakhondi2019}
Z.~Hajiakhondi-Meybodi, J.~Abouei, and A.~H.~F.~Raouf,
\newblock ``Cache Replacement Schemes Based on Adaptive Time Window for Video on Demand Services in Femtocell Networks,''
\newblock {\em IEEE Transactions on Mobile Computing}, vol. 18, no. 7, pp. 1476-1487, July 2019.

\bibitem{Hajiakhondi2020}
Z. HajiAkhondi-Meybodi, J. Abouei, M. Jaseemuddin and A. Mohammadi,
\newblock ``Mobility-Aware Femtocaching Algorithm in D2D Networks Based on Handover,''
\newblock {\em IEEE Trans. Veh. Technol.}, vol. 69, no. 9, pp. 10188-10201, June 2020.

\bibitem{Hajiakhondi2021-2}
Z. HajiAkhondi-Meybodi, A. Mohammadi, J. Abouei, M. Hou, and K. N. Plataniotis, \newblock ``Joint Transmission Scheme and Coded Content Placement in Cluster-centric UAV-aided Cellular Networks,''
\newblock {\em arXiv preprint arXiv:2101.11787}, July. 2021.

\bibitem{Giovanidis2016}
A. Giovanidis, and A. Avranas, \newblock ``Spatial multi-LRU caching for wireless networks with coverage overlaps,''
\newblock {\em ACM SIGMETRICS Performance Evaluation Review}, vol. 44, no. 1, pp. 403-405, June 2016.

\bibitem{Doan2018}
K. N. Doan, T. Van Nguyen, T. Q. S. Quek, and H. Shin, \newblock ``Content-Aware Proactive Caching for Backhaul Offloading in Cellular Network,''
\newblock {\em IEEE Trans. Wireless Commun.}, vol. 17, no. 5, pp. 3128-3140, May 2018.

\bibitem{Ale2019}
L. Ale, N. Zhang, H. Wu, D. Chen and T. Han, \newblock ``Online Proactive Caching in Mobile Edge Computing Using Bidirectional Deep Recurrent Neural Network,''
\newblock {\em IEEE Internet of Things Journal}, vol. 6, no. 3, pp. 5520-5530, June 2019.

\bibitem{Fan2021}
Q. Fan, X. Li, J. Li, Q. He, K. Wang and J. Wen, \newblock ``PA-Cache: Evolving Learning-Based Popularity- Aware Content Caching in Edge Networks,''
\newblock {\em IEEE Transactions on Network and Service Management}, vol. 18, no. 2, pp. 1746-1757, June 2021.

\bibitem{Zhang2019}
C. Zhang et al., \newblock ``Toward Edge-Assisted Video Content Intelligent Caching With Long Short-Term Memory Learning,''
\newblock {\em IEEE Access}, vol. 7, pp. 152832-152846, 2019.

\bibitem{Rathore2019}
S. Rathore, J. H. Ryu, P. K. Sharma and J. H. Park, \newblock ``DeepCachNet: A Proactive Caching Framework Based on Deep Learning in Cellular Networks,''
\newblock {\em IEEE Network}, vol. 33, no. 3, pp. 130-138, May/June 2019.

\bibitem{Lin2020}
Y. Lin, C. Yen and J. Wang, \newblock ``Video Popularity Prediction: An Autoencoder Approach With Clustering,''
\newblock {\em IEEE Access}, vol. 8, pp. 129285-129299, 2020.

\bibitem{Zhong2020}
C. Zhong, M. C. Gursoy and S. Velipasalar, \newblock ``Deep Reinforcement Learning-Based Edge Caching in Wireless Networks,''
\newblock {\em IEEE Trans. Cogn. Commun. Netw.}, vol. 6, no. 1, pp. 48-61, March 2020.

\bibitem{Wu2019}
P. Wu, J. Li, L. Shi, M. Ding, K. Cai and F. Yang, \newblock ``Dynamic Content Update for Wireless Edge Caching via Deep Reinforcement Learning,''
\newblock {\em IEEE Commun. Lett.}, vol. 23, no. 10, pp. 1773-1777, Oct. 2019.


\bibitem{Wang2019}
Y. Wang, Y. Li, T. Lan and V. Aggarwal, \newblock ``DeepChunk: Deep Q-Learning for Chunk-Based Caching in Wireless Data Processing Networks,''
\newblock {\em IEEE Trans. Cogn. Commun. Netw.}, vol. 5, no. 4, pp. 1034-1045, Dec. 2019.


\bibitem{Yu2021}
Z. Yu, J. Hu, G. Min, Z. Zhao, W. Miao and M. S. Hossain, \newblock ``Mobility-Aware Proactive Edge Caching for Connected Vehicles Using Federated Learning,''
\newblock {\em IEEE Trans. Intell. Transp. Syst.}, vol. 22, no. 8, pp. 5341-5351, Aug. 2021.

\bibitem{Tsai2018}
K. C. Tsai, L. Wang and Z. Han, \newblock ``Mobile Social Media Networks Caching with Convolutional Neural Network,''
\newblock {\em IEEE Wireless Communications and Networking Conference Workshops}, 2018, pp. 83-88.

\bibitem{Ndikumana2021}
A. Ndikumana, N. H. Tran, D. H. Kim, K. T. Kim and C. S. Hong, \newblock ``Deep Learning Based Caching for Self-Driving Cars in Multi-Access Edge Computing,'' \newblock {\em IEEE Trans. Intell. Transp. Syst.}, vol. 22, no. 5, pp. 2862-2877, May 2021.

\bibitem{Mou2019}
H. Mou, Y. Liu and L. Wang, \newblock ``LSTM for Mobility Based Content Popularity Prediction in Wireless Caching Networks,''
\newblock {\em IEEE Globecom Workshops}, 2019, pp. 1-6.



\bibitem{Vaswani2017}
A. Vaswani, N.Shazeer, N. Parmar, J. Uszkoreit, L. Jones, A. N. Gomez, L. Kaiser, and I. Polosukhin, \newblock ``Attention is all you need,''
\newblock {\em Advances in neural information processing systems}, pp. 5998-6008. 2017.

\bibitem{Zhang2021}
Y. Zhang, C. Li, T. H. Luan, C. Yuen, Y. Fu, H. Wang, H. and W. Wu, \newblock ``Towards Hit-Interruption Trade off in Vehicular Edge Caching: Algorithm and Analysis,''
\newblock {\em IEEE Trans. Intell. Transp. Syst.}, 2021.

\bibitem{Zhang2021_2}
Z. Zhang and M. Tao, \newblock ``Deep Learning for Wireless Coded Caching With Unknown and Time-Variant Content Popularity,''
\newblock {\em IEEE Trans. Wireless Commun.}, vol. 20, no. 2, pp. 1152-1163, Feb. 2021.



\bibitem{Chen2017_2}
Z. Chen, J. Lee, T. Q. S. Quek and M. Kountouris,
\newblock ``Cooperative Caching and Transmission Design in Cluster-Centric Small Cell Networks,''
\newblock {\em IEEE Trans. Wireless Commun.}, vol. 16, no. 5, pp. 3401-3415, May 2017.

\bibitem{Hajiakhondi2021}
Z. Hajiakhondi-Meybodi, A. Mohammadi and J. Abouei, \newblock ``Deep Reinforcement Learning for Trustworthy and Time-Varying Connection Scheduling in a Coupled UAV-Based Femtocaching Architecture,''
\newblock {\em IEEE Access}, vol. 9, pp. 32263-32281, Feb. 2021.


\bibitem{Wang2020}
X. Wang, C. Wang, X. Li, V. C. M. Leung and T. Taleb, \newblock ``Federated Deep Reinforcement Learning for Internet of Things With Decentralized Cooperative Edge Caching,''
\newblock {\em IEEE Internet of Things Journal}, vol. 7, no. 10, pp. 9441-9455, Oct. 2020.

\bibitem{Harper2015}
F. M. Harper,  and J. A. Konstan, \newblock ``The Movielens Datasets: History and Context,''
\newblock {\em ACM transactions on interactive intelligent systems}, vol. 5, no. 4, pp. 1-19, 2015.

\bibitem{Dernbach2016}
S. Dernbach, N. Taft, J. Kurose, U. Weinsberg, C. Diot and A. Ashkan, \newblock ``Cache Content-Selection Policies for Streaming Video Services,''
\newblock {\em IEEE International Conference on Computer Communications (INFOCOM)}, 2016, pp. 1-9.

\bibitem{Li2016}
S. Li, J. Xu, M. van der Schaar and W. Li, \newblock ``Popularity-Driven Content Caching,''
\newblock {\em IEEE International Conference on Computer Communications (INFOCOM)}, 2016, pp. 1-9.

\bibitem{Joseph2001}
C. Joseph, H. Hong, and J. C. Stein, \newblock ``Forecasting Crashes: Trading Volume, Past Returns, and Conditional Skewness in Stock Prices,''
\newblock {\em  Journal of Financial Economics}, vol. 61, no. 3, pp. 345-381, 2001.


\bibitem{Bert}
J. Devlin, M.W. Chang, K. Lee, and K. Toutanova,
\newblock ``Bert: Pre-training of Deep Bidirectional Transformers for Language Understanding,''
\newblock {\em arXiv preprint arXiv:1810.04805}, 2018.

\bibitem{layernorm}
JL Ba, JR Kiros, and G.E. Hinton,
\newblock ``Layer normalization,''
\newblock {\em arXiv preprint arXiv:1607.06450}, 2016.




\bibitem{Hong2020}
Y. Hong, J. J. F. Martinez, and A. C. Fajardo, \newblock ``Day-Ahead Solar Irradiation Forecasting Utilizing Gramian Angular Field and Convolutional Long Short-Term Memory,''
\newblock {\em  IEEE Access}, vol. 8, pp. 18741-18753, 2020.

\bibitem{Nguyen2019}
M. T. Nguyen, D. H. Le, T. Nakajima, M. Yoshimi and N. Thoai, \newblock ``Attention-Based Neural Network: A Novel Approach for Predicting the Popularity of Online Content,''
\newblock {\em  IEEE International Conference on High Performance Computing and Communications}, 2019, pp. 329-336.



\end{thebibliography}
\end{document}